
\input harvmac
\Title{\vbox{\baselineskip12pt\hbox{}\hbox{}}}
{\vbox{\centerline{Ordering, symbols, and finite-dimensional}
       \vskip3pt
       \centerline{ approximations of path integrals}}}

\centerline{Taro Kashiwa\footnote{$^*$}{E-mail address:
kashiwa@jpnyitp.bitnet,
}, Seiji Sakoda\footnote{$^{**}$}{E-mail address:
h79317a@kyu-cc.cc.kyushu-u.ac.jp}, Sergei V. Zenkin
\footnote{$^{***}$}{Permanent address: Institute for  Nuclear Research of the
Russian Academy of Sciences, Moscow 117312, Russia. E-mail address:
zenkin@inr.msk.su}}
\bigskip \centerline{Department of Physics, Kyushu University}
\centerline{Fukuoka 812, Japan}
\vskip 1.5cm

We derive general form of finite-dimensional approximations of path
integrals for both bosonic and fermionic canonical systems in terms of
symbols
of operators determined by operator ordering. We argue that for a system with
a given
quantum Hamiltonian such approximations are independent of the type of
symbols up to terms
of $O(\epsilon)$, where $\epsilon$ is infinitesimal time interval determining
the accuracy of
the approximations. A new class of such approximations is found for both
c-number
and Grassmannian dynamical variables. The actions determined by the
approximations are non-local and have no classical continuum limit except the
cases
of $pq$- and $qp$-ordering. As an explicit example the fermionic oscillator
is considered
in detail.
\Date{}

\newsec{Introduction}

Despite the fact that path integrals \ref\Fe{R. P. Feynman, Rev. Mod. Phys.
{\bf 20},
367 (1948);  Phys. Rev. {\bf 84}, 108 (1951)} have a long hisqory as a
standard technique in quantum physics, in particular in quantum field theory,
still
little is known about their exact meaning beyond those of the Gaussian
form. In the case of quantum systems involving both bosonic and fermionic
degrees of
freedom it is most natural to define path integrals as a limit of certain
multiple integrals when their multiplicity tends to infinity (for a review
and further
references see \ref\Be{F. A. Berezin, Sov. Phys. Usp. {\bf 23}, 763 (1980)}).
Such
multiple integrals considered to be finite-dimensional approximations
(FDA) of path integrals. In fact, it is also FDA that are a basis for the
lattice
approach to the constructive definition of quantum field theories, which
nowadays is
the most promising and powerful method to deal with them beyond  the
perturbation
theory.

It has been known since \ref\Ber{F. A. Berezin, Theor. Math. Phys. (USSR)
{\bf 6},
141 (1971)}  that the limit of FDA of path integrals over phase space depends
on the
choice of points in which  the exponent of FDA is calculated, the
ambiguity being closely related to that of operator ordering in canonical
quantization. Later, but independently and in another terminology, the issue
was
discussed extensively\nref\Mi{I. M. Mizrahi, J. Math. Phys. {\bf 16 }, 2201
(1975)}\nref\Co{L. Cohen,  J. Math. Phys. {\bf 17}, 597 (1976)}\nref\Do{ J.
S. Dowker, J.
Math. Phys.  {\bf 17}, 1873 (1976)}\nref\Sa{ M. Sato, Prog. Theor. Phys. {\bf
58}, 1262
(1976)}\nref\Ka{T. Kashiwa, Prog. Theor. Phys. {\bf 64}, 2164 (1980)} in \Mi
--\Ka\  (see also
\ref\Sc{L. S. Schulman, Techniques and Applications of Parth Integration  (A
Wiley-Interscience Publ., 1981), p. 257} for more references, and \ref\FS{L.
D. Faddeev, A. A.
Slavnov, Gauge Fields, Introduction to Quantum Theory, 2nd
edition (The Addison-Wesley Publ., 1990)}). The common conclusion was that
operator ordering
mainly affects the choice of the arguments of the Hamiltonian in the exponent
of FDA. This
leads to a naive expectation that the path integrals for a system with given
quantum
Hamiltonian is independent of the ordering, and that its exponent can always
be regarded as
the classical action of the system
$\int dt [ p \dot q - H(p, q) ]$ up to possible terms of
$O(\hbar)$ \Do. The former has been proven in perturbation theory in
\Sa\ and \Ka.  However, in \ref\Ze{S. V. Zenkin, Sov. Phys. Lebedev Inst.
Rep. {\bf 9},
10 (1988); Mod. Phys. Lett. {\bf A6}, 151 (1991)}, following the line of
\Ber\ and
\ref\BM{F. A. Berezin, M. S. Marinov, Ann. Phys. (N. Y.) {\bf 104}, 336
(1977)},
there has been found FDA for a fermionic system in the case of the Weyl
ordering
which appeared to be a counter example of the latter statement: its exponent
does not make
sense of integral sum of any local expression which could be considered
as a Lagrangean of the system.

In this paper we undertake an analysis of the the general structure of FDA
for both
bosonic and fermionic systems with the aim to ``dot the i's and cross the
t's" in this
issue. In particular, we demonstrate that for ``almost all" types of ordering
there
exist both representations of FDA with local and non-local forms in the
exponent, and
show in which sense path integrals are independent of ordering.

The outline and the results of the paper are as follows.

In Sect. 2 we give basic notations and technique which allows formal analogy
between
bosonic and fermionic quantum systems. This is achieved by using eigenvectors
of
operators of canonical momentum and coordinate for the bosonic system, and a
coherent state like representation for the fermionic system \ref\OK{Y.
Ohnuki, T.
Kashiwa, Prog. Theor. Phys. {\bf 60}, 548 (1978)}.

In Sect.3 we introduce the notion of symbols of operators: functions defined
on the
phase space which are in certain one-to-one correspondence with
operators on the Hilbert space. There are various such mappings, each one
being uniquely determined by the operator ordering. Our definitions
generalize those
of \ref\Coh{L. Cohen, J. Math. Phys. {\bf 7}, 781 (1966)} and of \Ber, \BM\
to
Grassmannian variables and to arbitrary ordering, respectively.
We consider general ordering (and corresponding symbols) determined by
an analytical function and called $f$-ordering ($f$-symbols). Then we
emphasize a
special role of particular case of $f$-ordering, the so called
$\alpha$-ordering \Ka. In
this case the function is parameterized by a single parameter $\alpha$, so
that the
Weyl, ${p} {q}$ and
$ {q} {p}$ orderings correspond to $\alpha = 0$, $1/2$, and $-1/2$,
respectively. This scheme is still the most general for fermionic systems.

In Sect. 4  we construct FDA for both bosonic and fermionic systems in
terms of the $f$-symbols.  Our construction uses standard ``resolution of
unity'',
rather than $*$-products determining representations of the operator algebra
in
terms of their symbols which have been used in \Ber, \BM, and \Ze. Therefore,
we obtain
the local form of FDA found earlier in \Do\ and \Ka. A new point here is
arguments
in favour that FDA for a system with a given quantum Hamiltonian are
independent of type of the
symbols up to terms
$O(\epsilon)$, where
$\epsilon$ is (infinitesimal) time interval determining the accuracy of the
FDA.

In Sect. 5 we consider FDA for both bosonic and fermionic systems in
terms of $\alpha$-symbols in more detail.  In this case kinetic part of the
FDA
corresponding to classical term $p \dot q$ is represented by the finite
difference
$p_i (q_i - q_{i-1})$, while the Hamiltonian part is taken in the points
$p_i$ and
$q^{(\alpha)}_i = ({1 \over 2} - \alpha) q_i + ({1 \over 2} + \alpha)
q_{i-1}$, being
in this sense weekly non-local. Then making a change of variables we find the
new
representation for the exponent of FDA: except
$\alpha = \pm 1/2$ the kinetic part of the exponent is highly non-local and
has no
naive continuum limit, while the Hamiltonian part is strictly local, that is
taken
in the points $p_i$, $q_i$. The situation is the same in the fermionic case.
This
results generalize to any $\alpha$ as well as to bosonic case the one of \Ze.
We conclude this section with explicit demonstration of $\alpha$-independence
up
to $O(\epsilon)$ of FDA for the fermionic oscillator.

Sect. 6 is a short summary and discussion.

For the sake of notation convenience we make all the explicit constructions
for a
system with a single, either bosonic or fermionic, dynamical variable. The
generalization to any finite number of variables is straightforward.

\newsec{Basic notations and definitions}

We denote all operators with hats, so that notation $\hat H(\hat p, \hat q)$
means
operator function $H(\hat p, \hat q)$.

\subsec{ Bosonic system (c-numbered dynamical variable)}

Quantum bosonic system is defined by commutation relations
\eqn\Iaa{ [{\hat q}, {\hat p}] = i, \quad  [{\hat q},
{\hat q}] = [{\hat p},{\hat p}] =0,}
with $\hat p$ and $\hat q$ being operators of canonical momentum
and coordinate, respectively,  acting on the Hilbert space ${\cal H} = {\cal
L}_2$ (we
put
$\hbar = 1$). Operators $\hat p$ and $\hat q$ are self-adjoint on rigged
Hilbert space
${\bar{\cal H}} = {\cal S} \subset {\cal H} \subset {\cal S}^*$
\ref\BLT{N. N. Bogolubov, A. A. Logunov, I. T. Todorov, Introduction to
Axiomatic
Quantum Field Theory (W. A. Benjamin, 1975)}. Let $|p\rangle$, $|q\rangle
\in {\bar{\cal H}}$ be complete sets of eigenvectors of the operators with
eigenvalues
$p \in (-\infty, \infty)$ and
$q \in (-\infty, \infty)$:
\eqn\pqa{\eqalign{
& \hat q\left| q \right\rangle =q\left| q \right\rangle ,\quad \hat p\left| p
\right\rangle =p\left| p \right\rangle, \cr
& \left\langle q \right|\hat q =\left\langle q \right| q,\quad
\left\langle p \right|\hat p =\left\langle p \right| p. \cr }}
Then we have
\eqn\pqb{\eqalign{
& \left\langle {q} \mathrel{\left | {\vphantom {q p}} \right.
\kern-\nulldelimiterspace} {p} \right\rangle ={1 \over \sqrt{2\pi} }
\e{ipq},\quad
\left\langle {p} \mathrel{\left | {\vphantom {p q}} \right.
\kern-\nulldelimiterspace} {q} \right\rangle ={1 \over \sqrt{2\pi} }
\e{-ipq},
\cr & \left\langle {q} \mathrel{\left | {\vphantom {q {q'}}} \right.
\kern-\nulldelimiterspace} {{q'}} \right\rangle =\delta \left( {q-q'}
\right),\quad \left\langle {p} \mathrel{\left | {\vphantom {p {p'}}} \right.
\kern-\nulldelimiterspace} {{p'}} \right\rangle =\delta \left( {p-p'}
\right); \cr }}
\eqn\pqc{
\int {dq}\left| q \right\rangle \left\langle q \right|={\hat
I},\quad
\int{dp}\left| p\right\rangle \left\langle p \right|={\hat I}; }
\eqn\pqd{
\Tr \hat O = \int {dq} \left\langle q \right| \hat O \left| q \right\rangle,
}
where $\hat O$ is an operator on $\bar{\cal H}$.

\subsec{Fermionic system (Grassmannian dynamical variable)}

Here we give a brief sketch of the formalism developed in \OK\ which will
allow us to construct
path integrals for the fermionic systems using, in fact, the same technique
as for
bosonic systems.

The simplest fermionic system is defined by the anticommutation relations
\eqn\ope{\{\hat b, \hat b^{\dag} \} = 1, \quad \{\hat b, \hat b\} =
\{\hat b^{\dag}, \hat b^{\dag}\} = 0,}
with $\hat b^{\dag}$ and $\hat b$ being operators conjugate with respect to
each
other on a Hilbert space
${\cal H}$. By
virtue of nilpotency of $\hat b^{\dag}$ and $\hat b$ the Hilbert space is two
dimensional: let $|0\rangle$ and $|1 \rangle$ be basic orthonormal
vectors of $\cal H$, then $\hat b|0\rangle = 0$, $\hat b^{\dag}|0\rangle =
|1\rangle$.
Vectors of $\cal H$ are elements of a Grassmann algebra with inner product
\ref\Bere{F. A. Berezin, The Method of Second Quantization (Academic Press,
1966)}. We use the convention that $|0\rangle$ ($|1\rangle$) is even (odd)
element of the algebra.

Define the space ${\cal B} \subset {\cal H} \times {\cal G}$, where ${\cal
G}$ is a
Grassmann algebra with involution and generators $\xi_i$, such that
$|\psi\rangle =
\psi_0(\xi) |0\rangle + \psi_1(\xi) |1\rangle \in {\cal B}$,  $\psi(\xi) =
c_0 + \sum_i
c_i \xi_i$, $c_{0, i}$ are complex numbers.  It is convenient to define also
conjugate
space ${\cal B}^*$ with elements $\langle\psi'^*| = \langle0|\psi'^*_0 (\xi)
+ \langle1|
\psi'^*_1 (\xi)$ such that
\eqn\bb{\langle\psi'^*| \psi\rangle = \psi'^*_0(\xi) \psi_0(\xi) +
\psi'^*_1(-\xi)
\psi_1(-\xi) \in {\cal G},}
where star for Grassmann numbers means involution. Thus product \bb\ is
Hermitean
in the sense of the involution: $\langle\psi'^*| \psi\rangle^* =
\langle\psi^*|
\psi'\rangle$.

We use the usual conventions for differentiation and integration on ${\cal
G}$
\Bere, except the following one:
\eqn\integ{\int \xi d\xi = i}
($\xi$ is some of $\xi_i$), so that $\delta$-function /n $\cal G$ is
\eqn\del{\delta(\xi - \xi') \equiv {1 \over i} (\xi - \xi') = \int \e{-(\xi -
\xi')\xi''} d
\xi''.}
In ${\cal B}$ and ${\cal B}^*$ there exist vectors
\eqn\cs{\eqalign{& |\xi\rangle \equiv \e{-\xi \hat b^{\dag}} |0\rangle =
|0\rangle -
\xi |1\rangle, \quad  |\xi^*\rangle \equiv \delta (\hat b^{\dag} - \xi^*)  =
i
\xi^*|0\rangle - i |1\rangle, \cr
& \langle \xi| \equiv \langle0| \delta(\xi - \hat b) = - \langle0| i \xi + i
\langle1|,
\quad \langle \xi^*| \equiv \langle0| \e{\xi^* \hat b} = \langle0| -
\langle1| \xi^*,}}
which are eigenvectors of operators $\hat b$ and $\hat b^\dagger$,
respectively:
\eqn\cspr{\eqalign{& \hat b |\xi\rangle = \xi |\xi\rangle, \quad  \hat
b^{\dag}
|\xi^*\rangle =
\xi^* |\xi^*\rangle,\cr
& \langle \xi| \hat b = \langle \xi| \xi, \quad  \langle
\xi^*| \hat b^{\dag} = \langle \xi^*| \xi^*. \cr }}
Then it immediately follows from \bb\ -- \cspr\ that
\eqn\eq{
\eqalign{
&\langle \xi| \xi'^* \rangle = \e{\xi \xi'^*}, \quad \langle \xi^*| \xi'
\rangle = \e{\xi^*\xi'}, \cr
& \langle \xi'| \xi \rangle = \delta(\xi'- \xi), \quad \langle \xi'^*| \xi^*
\rangle = \delta(\xi'^* - \xi^*); \cr
}}
\eqn\eqa{\eqalign{
& \int |\xi \rangle \langle \xi| d \xi = \int d \xi^* |\xi^* \rangle
\langle \xi^*| = |0\rangle \langle0| + |1\rangle \langle1| = \hat I; \cr
}}
\eqn\tr{
Tr \hat \Omega = \int \langle \xi| \hat \Omega |-\xi\rangle d \xi,
}
where $\hat \Omega$ is a Grassmann even operator on $\cal H$. For a proof of
such
relations in the case of any finite number degrees of freedom (i.e. pairs of
operators
$\hat b^\dagger_i$, $\hat b_i$) see \OK.

Comparing \pqa, \pqb, and \pqc\ with  \cspr, \eq, and \eqa, we notice a
formal
analogy between vectors $|\xi\rangle$, $|\xi^*\rangle$, and $|q\rangle$,
$|p\rangle$,
respectively. This analogy becomes more substantial if we note that operators
$\hat b$
and $- i \hat b^\dagger$ are actually operators of canonical coordinate and
conjugate
momentum of the fermionic system \ref\Zen{S. V. Zenkin, Lattice fermions and
the
structure of gauge theories, preprint of Pisa University IFUP-TH 45/90
(1990)}.

\newsec{Operators and their symbols}

Subject of this section is correspondence between functions of the form
$A(p, q) =\sum_{m, n \geq 0} A_{mn} p^m q^n$ on the phase space,
or $A(\xi^*, \xi) =
\sum_{m, n = 0, 1} A_{mn} {\xi^*}^m \xi^n$ on the Grassmann algebra $\cal G$,
and
operators $\hat A(\hat p, \hat q)$, or $\hat  A(\hat b^\dagger, \hat b)$, on
the
corresponding Hilbert spaces. Namely, we shall set up various one-to-one
mappings
between the functions and operators.

\subsec{Bosonic system}

Let $f(u, v)$ be a function analytical with all its derivatives in some
vicinity of the
origin
$D_{\varepsilon}$ such that $Re f > 0$ in $D_{\varepsilon}$, and satisfying
condition
\eqn\IIc{
 f(u,0) = f(0,v) = 1.
}
Define {\it f-ordered} operator $\hat A^{(f)} ({\hat p},{\hat q})$
corresponding to
function $A(p, q)$ as \Coh
\eqn\IIa{
\hat A^{(f)} ({\hat p},{\hat q}) \equiv \int f(u,v) \e{-i({\hat p}-p)v
-i({\hat
q}-q)u} A(p,q) {dp\over 2\pi} dq {du\over 2\pi} dv.
}
So, \IIa\ defines a mapping:
\eqn\IIb{f: \quad A(p,q) \quad {\longmapsto} \quad \hat A^{(f)} ({\hat
p},{\hat
q}).}
Condition \IIc\ provides a natural requirement to the mapping: $p^m \mapsto
{\hat p}^m$, $q^n \mapsto {\hat q}^n$. If function $f$ satisfies condition
\eqn\real{
f^*(u, v) = f(-u, -v),
}
operator $\hat A^{(f)} ({\hat p},{\hat q})$ is  Hermitean, providing  $A(p,
q)$ is real.

Let us now show that any operator $\hat A(\hat p, \hat q)$ can be reduced
to $f$-ordered form with arbitrary $f$ satisfying the above conditions.

Introduce a generating function of $f$-ordered operators
\eqn\gene{
\eqalign{\hat G^{(f)}\left(\hat{p},\hat{q};a,b\right)&\equiv\int f(u,v)
\e{-i({\hat
p}-p)v -i({\hat q}-q)u} \e{-iap-ibq} {dp\over 2\pi} dq {du\over 2\pi} dv\cr
& = f(b, a) \e{-ia\hat{p}-ib\hat{q}}, \cr
} }
so that
\eqn\IIf{\eqalign
{
\{{\hat p}^m{\hat q}^n \}^{(f)} &\equiv \int f(u,v) \e{-i({\hat p}-p)v
-i({\hat q}-q)u}
 p^{m}q^{n} {dp\over 2\pi} dq {du\over 2\pi} dv \cr &=\left. {\left(
{i{\partial  \over
{\partial a}}} \right)^m\left( {i{\partial  \over {\partial b}}}
\right)^nG^{\left( f
\right)}\left( {\hat p,\hat q;a,b} \right)}
\right|_{a=b=0}.\cr
} }
Then, since the generating functions according to \gene\ are related to each
other as
\eqn\trans{
 G^{\left( f_{1} \right)}\left( {\hat p,\hat q;a,b}
\right)=\left[{{f_{1}(b, a)}\over{f_{2}(b, a)}}\right]G^{\left( f_{2}
\right)}\left(
{\hat p,\hat q;a,b} \right),\quad a,b \in D_\varepsilon,
}
we have
\eqn\gtrans
{
\{{\hat p}^m{\hat q}^n
\}^{(f_{1})}=\sum_{k=0}^{m}\sum_{l=0}^{n} T^{m n}_{k l}(f_{1},f_{2})
\{{\hat p}^k{\hat q}^l \}^{(f_{2})},
}
where coefficients $T^{m n}_{k l}(f_{1},f_{2})$ determined by the Leibnitz
rule have
the form
\eqn\coef{
T^{m n}_{k l}(f_{1},f_{2})\equiv
\left. {\left( {\matrix{m\cr k\cr }} \right) \left( {\matrix{n\cr l\cr }}
\right) \left( {i{\partial  \over {\partial a}}} \right)^{m-k}\left(
{i{\partial
\over {\partial b}}} \right)^{n-l}\left[ {{{f_{1}\left( {b, a} \right)} \over
{f_{2}\left(
{b, a} \right)}}} \right]} \right|_{a=b=0}.
}
In terms of complex indices $M, K = (m, n), \ldots, (m, 0), (m-1, n), \ldots
(m-1, 0), \ldots (0, n)$, $\ldots (0, 0)$ (so that $M$ and $K$ take $(m + 1)
(n + 1)$
values) matrix $T^M_K$ is always triangle with diagonal elements $T^K_K = 1$.
Therefore any polynomial of the form $\{{\hat p}^m{\hat q}^n \}^{(._{1})}$
can be
reordered according to another ordering function $f_2$, transformation
\gtrans\ being
regular for any $f_1$, $f_2$.

To complete our proof, note that any operator of polynomial form by
subsequent
commutations of operators $\hat p$ and $\hat q$ in each its monomial can be
reduced
to the $pq$-ordered form
\eqn\pqord{
\hat A(\hat p, \hat q) = \sum_{m, n} A^{(pq)}_{m n} \hat p^m \hat
q^n.}
Then, it immediately follows that $\hat A(\hat p, \hat q)$ can be reduces
to arbitrary  $f$-ordered form
\eqn\ford{
\hat A(\hat p, \hat q) = \sum_{mn} A^{(f)}_{m n}\{\hat p^m \hat
q^n\}^{(f)},
}
where, according to \gtrans,
\eqn\pqf{
A^{(f)}_{m n} = \sum_{k, l} A^{(pq)}_{k l} T^{k l}_{m n} (pq, f).
}

Define a function on the phase space $A^{(f)}(p, q)$ called {\it
f-symbol} of operator $\hat A(\hat p, \hat q)$ as
\eqn\fsymb{
A^{(f)}(p, q) = \sum_{mn} A^{(f)}_{m n} p^m q^n.
}
Eq. \fsymb\ in view of \ford\ defines a mapping inverse to \IIb:
\eqn\opsymb{
f^{-1}: \quad \hat A({\hat p},{\hat q})\quad {\longmapsto} \quad
A^{(f)}(p,q).
}

By noting that the $pq$-symbol of any operator reads as
\eqn\IIk{
 A^{(pq)}(p,q) = {{\langle{p |{ \hat A ({\hat p},{\hat
q})}|q}\rangle} \over {\langle p|q\rangle}},
}
from \IIa, \ford, and \fsymb\ we find
\eqn\pqsym{
\eqalign{ A^{(pq)}(p,q) &= \int\tilde{f}(u,v) \e{-i(p-p')v -i(q-q')u}
A^{(f)}(p',q')
{dp'\over 2\pi} dq' {du\over 2\pi} dv \cr

& = \tilde{f} ( i{\partial \over \partial q} , i{\partial \over \partial p})
A^{(f)}(p, q),
\cr}}
where we denote
\eqn\ftilde{
\tilde{f} (u,v)\equiv f(u,v) \e{-{i\over 2}uv }.
}

{}From the same formulae we find representation of the matrix elements
$\langle q_2|\hat A({\hat p},{\hat q}) |q_1\rangle$ of the arbitrary operator
$\hat
A({\hat p},{\hat q})$ in terms of its symbols:
\eqn\IIla{
\langle q_2|\hat A({\hat p},{\hat q}) |q_1\rangle = \int {dp\over {2\pi}}
\e{ip(q_2
-q_1)}  A^{(f)}\left({p, q_2, q_1}\right), }
where
\eqn\IIlb{
\eqalign{ A^{(f)}\left({p, q_2, q_1}\right) &
\equiv \tilde{f} ( i{\partial \over \partial q} , q_2-q_1)
A^{(f)}(p,q)\Bigg|_{q=q_1}.
\cr
}}
Expressions \IIla\ and \IIlb\ play a central role in our construction of path
integrals in
Sect.4.

There is an important particular case of $f$-ordering in which
the function $f$ is parameterized by a single real parameter $\alpha$:
\eqn\IIm{
 f(u,v) = \e{i\alpha uv}.
}
We shall call it {\it $\alpha$-ordering} \Ka. Most
of ordering schemes so far discussed are either special cases of
$\alpha$-ordering, or
linear combinations of them with different $\alpha$'s. For example:

$pq$-ordering: $ \quad \alpha = {1 \over 2}, \quad \{\hat p^m \hat
q^n\}^{(pq)} =
\hat p^m \hat q^n;$

$qp$-ordering: $\quad \alpha = -{1 \over 2}, \quad \{\hat p^m \hat
q^n\}^{(qp)} =
\hat q^n \hat p^m ;$

Weyl ordering: $\quad \alpha = 0, \quad \{\hat p^m \hat q^n\}^{(W)} =
2^{-n}
\sum_{l = 0}^{n} { n\choose l}  \hat q^{n - l} \hat p^m \hat q^l $  \Coh;

Symmetric ordering: $f(u, v) = \cos{1 \over 2}u v,\quad \{\hat p^m
\hat q^n\}^{(sym)} = {1 \over 2}(\hat q^n \hat p^m + \hat p^m \hat q^n). $

A special role of $\alpha$-ordering  in construction of path integrals is
determined
by the fact that function $A^{(f)}({p, q_2, q_1})$ in \IIlb\ in this case
takes
particularly simple form:
\eqn\IIp{
\eqalign{ A^{(\alpha)}(p, q_2, q_1)  =  & \e{({1\over 2}
-\alpha)(q_2 - q_1) {\partial / \partial q_1} } A^{(\alpha)}(p,q_1) \cr
= & A^{(\alpha)}(p, q^{(\alpha)}(q_2, q_1)),\cr} }
where
\eqn\qalha{ q^{(\alpha)}(q_2,q_1)\equiv\left({{1\over 2} -\alpha}\right)q_2
+\left({
{1\over 2} +\alpha}\right)q_1. }

\subsec{\it Fermionic system}

By virtue of the nilpotency of both operators $\hat b^\dagger$, $\hat b$ and
generators
$\xi^*$, $\xi$ of
$\cal G$ the most general ordering in this case is just the
$\alpha$-ordering. We define the $\alpha$-ordered operator $\hat A^{(\ulpha)}
({\hat
b^\dagger},{\hat b})$ corresponding to function $A(\xi^*, \xi)$ as
\eqn\IIaf{
 \hat A^{(\alpha)} ({\hat b^\dagger},{\hat b}) \equiv \int \e{\alpha \zeta^*
\zeta}
\e{-i({\hat b^\dagger}-\xi^*)\zeta -i\zeta^*({\hat b}-\xi)} A(\xi^*, \xi)
d\xi d\xi^*
d\zeta d\zeta^*,
}
where $\zeta^*$ and $\zeta$ are Grassmann analogues of $u$ and $v$,
respectively.
Eq. \IIaf\ defines non-trivial mapping
\eqn\IIbf{[\alpha]: \quad \xi^* \xi \quad {\longmapsto} \quad ({1 \over 2} +
\alpha)\hat
b^\dagger \hat b - ({1 \over 2} - \alpha) \hat b \hat b^\dagger,}
in addition to: $ \xi^* \mapsto \hat b^\dagger$, $\xi \mapsto \hat b$.
Note, that contrary to the bosonic case, now there is no restriction for
$\alpha$ to
be real. In view of the above mentioned analogy  between variables and
operators of the fermionic and bosonic systems the cases
$\alpha = {1 \over 2}$, $0$, and $-{1 \over 2}$ directly correspond to $pq$
(normal),
Weyl, and $qp$ (antinormal) orderings, respectively.

$\alpha$-symbol $A^{(\alpha)}(\xi^*, \xi)$ of an operator $\hat A ({\hat
b^\dagger},{\hat b})$ is defined in the same way as in the bosonic case, so
that, for
example, we have:
\eqn\IIbff{[\alpha]^{-1}: \quad \hat b^\dagger \hat b
\quad {\longmapsto} \quad \xi^* \xi + {1 \over 2} - \alpha.}

Using the technique of Sect.2.2, in complete analogy with Eqs.\IIla, \IIp,
and \qalha,
we find
\eqn\ferr{
\eqalign{
\langle \xi_{2}  |A( {\hat b^\dagger ,\hat b} )| {\xi_{1}}
\rangle &= -\int d\xi^{*} {\e{-\xi ^*\left( {\xi_{2} -\xi_{1}} \right)}A^{(
\alpha  )}\left( {\xi ^*,\xi ^{( \alpha  )}(\xi_{2},\xi_{1})} \right)},\cr}}
where
\eqn\xialpha{\xi ^{\left( \alpha  \right)}(\xi_{2},\xi_{1}) \equiv \left( {{1
\over
2}-\alpha } \right)\xi_{2} +\left( {{1 \over 2}+\alpha } \right)\xi_{1}.}

Note that in both bosonic and fermionic cases the symbols
of operators form linear representations of the corresponding algebras of the
operators with operation of multiplication ($*$-product) of the symbols
defined by
the ordering functions $f$ (see \Ber, \Zen ).

\newsec{Finite dimensional approximations of path integrals}

In this section we construct finite-dimensional approximations (FDA) of path
integrals
for the Feynman kernel and for the partition function in terms of
$f$-symbols and ascertain some of the properties of the FDA.

Dynamics of a system is determined by Hamiltonian $\hat H$ which is a
function of
operators $\hat p$ and $\hat q$ ($\hat b^\dagger$ and $\hat b$ in the
fermionic case),
or, equivalently, by the evolution operator ${\hat U}(t, t_0)$ which
satisfies the
Schr\"odinger equation
\eqn\schr{ i{ \partial \over \partial t} {\hat U}(t, t_0) = {\hat H}(t){\hat
U}(t, t_0),
\quad {\hat U}(t_0, t_0) = {\hat  I},}
where for the sake of generality we allow a time dependence of the
Hamiltonian.

Our starting point is representation of the evolution operator in terms of
product
\eqn\consa{
\hat U( {t,t_0} ) = \lim_{N \rightarrow \infty} \hat U\left(
{t_N,t_{N-1}} \right) \hat U\left( {t_{N-1},t_{N-2}} \right)\cdots \hat
U\left( {t_2,t_1}
\right) \hat U\left( {t_1,t_0} \right)\quad,
}
where
\eqn\consb{
\hat U( {t_i,t_{i-1}} ) \equiv {\hat I} - i\epsilon \hat H(t_i), \quad
\epsilon = t_i - t_{i-1} ={{t-t_0}
\over N},\quad i = 1, \ldots, N, \quad t_N=t.}

Consider first

\subsec{Bosonic system}

The Feynman kernel is given by
\eqn\consc{
\langle q_F | \hat U(t,t_0)| q_I \rangle = \lim_{N
\rightarrow \infty}\int\prod\limits_{ j\ =\ 1}^{ N-1}  d{q}_{j}\prod\limits_{
i\ =\ 1}^{
N} \langle q_i | \hat U(t_i, t_{i-1}) |q_{i-1}\rangle,\  q_0=q_I,\ q_N=q_F.
}
Applying \IIla, \IIlb\ to each term of the product in \consc, with
\eqn\hammat{
\langle q_i | {\hat H}(t_i) |q_{i-1}\rangle  =  \int { d{p}_i \over 2\pi }
\e{ ip_i (q_i -
q_{i-1} )} {H}^{(f )}_{i}( p_i , q_i, q_{i-1}),
}
we have
\eqn\consd
{\eqalign {
\langle q_i | & \hat U(t_i,t_{i-1}) | q_{i-1} \rangle \cr = & \int { d{p}_{i}
\over
2\pi } \e{ ip_i (q_i-q_{i-1})} [1 - i\epsilon H^{(f)}_i( p_i , q_i, q_{i-1})]
 \cr =
&
\int { d{p}_{i}
\over 2\pi } \e{ ip_i (q_i-q_{i-1}) - i\epsilon H^{(f)}_i( p_i ,q_i,
q_{i-1})} +
O(\epsilon^2).
\cr
}}
We should note that actually the exponentiating in Eq.\consd\ is correct if
both the integral
and term $O(\epsilon^2)$ there are well-defined in the sense of generalized
functions
(distributions) (see, e.g., \BLT), so that $\lim_{N \rightarrow
\infty} [O(N) O(\epsilon^2)] = 0$ is hold at least in the same sense. In
physical
terminology this means that if damping factors such as $\e{-\kappa {p_i}^2}$
are inserted into
both expressions of \consd, the difference between them, being a well-defined
function of
$\epsilon$ and $\kappa$ at $\kappa \neq 0$, tends to zero at $\epsilon
\rightarrow 0$ as
$\epsilon^2$. Then limit $\kappa_1 \rightarrow 0$ results in the
$O(\epsilon^2)$ as a
generalized function. The conditions for that are closely related to those
for the form
of the Hamiltonian for which the evolution operator exists itself. We however
shall not
investigate these conditions and in what follows merely assume that they are
fulfilled.

Then from \consc\ and \consd\ we get the path integral representation for the
kernel
\eqn\conse{
\langle q_F | \hat U(t,t_0)| q_I \rangle = \lim_{N \rightarrow \infty}
U^{(f)}_N(t,t_0 ; q_F, q_I),
}
where the multiple integral
\eqn\consf{\eqalign{
U^{(f)}_N( & {t,t_0}; q_F, q_I) \cr
& = \int_{}^{} \prod_{ j\ =\ 1}^{ N-1}
d{q}_{j}\prod_{ i\ =\ 1}^{ N} {d{p}_{i} \over 2\pi}
\e{i[p_{i}\Delta q_{i} - \epsilon {H}_i^{({f })}( p_i , q_i, q_{i-1})]}, \cr
& q_0 = q_I,\; q_N = q_F, \quad \Delta q_{i} \equiv q_{i}-q_{i-1},\cr}
}
is called FDA of the path integral \conse\ in terms of $f$-symbols.

{}From the above construction it follows that dependence of FDA \consff on
type of the
symbols in terms of which it is formulated, i. e. on function $f$, is of
order
$O(\epsilon)$, while the path integral \conse\ is independent of $f$.

We can convince ourselves of this by another way. Let us suppose that FDA is
expressed in terms
of the $pq$-symbols, that is, according to
\IIp, \qalha\ its exponent reads as
\eqn\pqexp{
 i p_i\Delta q_i  - i \epsilon H^{(pq)}_i( p_i ,  q_{i-1}) ,
}
and show that it is equal to exponent of \consf, providing the neglection of
terms
$O(\epsilon^2)$. Let function
$\tilde f(u, v)$ determining \consf\ has an
expansion
\eqn\fexp{
\tilde f(u, v) = \sum_{l,m \geq 0} \tilde f_{lm} u^l v^m
 } so that the exponent in \consf\ is
\eqn\expo{
 i[p_i- \epsilon \sum_{l, m \geq 0} \tilde f_{lm} (\Delta q_i)^{m-1} (
i{\partial \over
\partial q_{i-1}} )^l H^{( f )}_i ( p_i,q_{i-1}) ] \Delta q_i .
}
Make now change of variable
\eqn\newp{
 p'_i = p_i - \epsilon \sum_{l \geq 0, m \geq 1}  \tilde f_{lm} ( \Delta q_i
)^{m-1}  ( i
{\partial  \over {\partial q_{i-1}}} )^l H^{( f)}_i( p_i,q_{i-1}).
}
Then its Jacobian
\eqn\jacobi{\eqalign{
{dp_i\over {dp'_i}}   =  & 1 - i \epsilon  \sum_{l \geq 0, m \geq 1}
\tilde f_{lm} (\Delta q_i)^{m-1}  i {\partial  \over {\partial p'_i }}  ( i
{\partial  \over
{\partial q_{i-1}}})^l H^{( f)}_i( p'_i,q_{i-1}) + O(\epsilon^2) \cr  = &
\exp[ -i \epsilon
\sum_{l \geq 0, m \geq 1} \tilde f_{lm} (\Delta q_i)^{m-1}  i {\partial
\over {\partial
p'_i }}  ( i {\partial  \over {\partial q_{i-1}}})^l H^{( f)}_i(
p'_i,q_{i-1}) ]  +
O(\epsilon^2) \cr
}}
yields the new exponent:
\eqn\consea{
i p_i \Delta q_i  - i \epsilon  \sum_{l \geq 0} [\tilde f_{l 0} + \sum_{
m\geq 1}
\tilde f_{lm} (\Delta q_i )^{m-1}  i{\partial  \over \partial p_i }  (i
{\partial  \over
\partial q_{i-1}} )^l ] H^{( f )}_i( p_i, q_{i-1}),
}
where we have omitted the prime in $p_i$. Following the same procedure
subsequently and
neglecting terms $O(\epsilon^2)$ we come to the ``substitution rule"
\ref\HFTK{H. Fukutaka and T. Kashiwa, Ann. Phys. {\bf 185}, 3011(1988)}
\eqn\substi{
\Delta q_i \quad \longrightarrow \quad i{\partial\over \partial p_i},
}
for the function $\tilde f(i \partial  / {\partial q}, \Delta q_i)$ in
Eq.\expo:
\eqn\conseb{
i p_i\Delta q_i  - i \epsilon \sum_{l,m \geq 0} \tilde
f_{lm} (i {\partial  \over
\partial q_{i-1} } )^l (i{\partial \over \partial p_i})^m H^{( f )}_i( p_i,
q_{i-1}).
}
By virtue of \pqsym\  it is exactly the exponent of FDA in terms of
$pq$-symbols \pqexp.

Using the trace formula \pqd\ from \consc\ and \consf\ we get expression for
FDA of path
integral for the trace of the evolution operator (``partition function")
\eqn\consg{\eqalign{  Z^{(f)}_N({t,t_0}) & = \int dq U^{(f)}_N({t,t_0}; q,
q)\cr
& = \int_{}^{}\prod_{ i\ =\ 1}^{ N}  d{q}_{i}{d{p}_{i} \over
2\pi} \;
\e{i [p_{i}\Delta q_{i} - \epsilon {H}_i^{({f })}( p_i , q_i, q_{i-1})]},
\quad q_0 = q_N.\cr} }

To conclude the discussion of the general form of FDA, note that,
in general, $f$-symbols of a Hamiltonian contain terms $O(\hbar)$. Given
quantum Hamiltonian
$\hat H(\hat p,\hat q)$, we can define classical Hamiltonian
$H_{cl}(p, q)$ as a limit at ${\hbar \rightarrow 0}$ of $H^{(f)}(p, q)$; the
limit is
independent of
$f$. Then, the exponent in FDA \consf\ is an integral sum for the classical
action of the
system $\int^{t}_{t_0} dt [p \dot{q} - H_{cl}(p, q)]$ only for $f$ or
$\hat H$ being of particular forms. For example, for $\alpha$-ordering this
is
the case if $\hat H$ is $\alpha$-ordered image (3.2) of the $H_{cl}$, or if
$\hat
H$ is of the form $\hat H(\hat p,\hat q) = \hat H_1(\hat p) + \hat H_2(\hat
q)$. In general, function ${H}^{\left({\alpha}\right)}$ can be complex even
if the
Hamiltonian is self-adjoint.

In the case of any finite number $M > 1$ degrees of freedom quantities $p_i$
and $q_i$ in
Eqs.\consf, \consg\ should be considered as $M$-dimensional vectors, so that,
i.e.,
$p_i q_j = \sum_{r = 1}^{M} p^r_i q^r_j$, $d p_i/(2 \pi) = \prod_{r = 1}^{M}
[d p^r_i/(2
\pi)]$, etc.

\subsec{Fermionic system}

As the kernel of the evolution operator makes no physical meaning in this
case, we
consider partition function $Z(t, t_0) = \Tr \hat U(t, t_0)$. The technique
of
Sect.2.2, as well as formulae of Sect.3.2, allow us to make all the
construction in a complete
analogy with the bosonic system. Then, considering the Hamiltonian to be an
even element of a
Grassmann algebra, we find
\eqn\fz{Z(t, t_0) = \lim_{N \rightarrow \infty} Z^{(\alpha)}_N(t, t_0),}
where
\eqn\fconsa{
\eqalign{Z^{(\alpha)}_N&(t, t_0) \cr
& = \int \prod\limits_{i=1}^{N} d\xi_i d\xi^*_i \e{- \xi _i^* (\xi _i -\xi
_{i-1}) - i \epsilon H_i^{(\alpha  )}(\xi_i^*, \xi^{(\alpha)}(\xi_i,
\xi_{i-1}))}, \quad
\xi_0 = - \xi_N, \cr} }
is FDA of the fermionic path integral \fz\ in terms of $\alpha$-symbols.
Antiperiodic
boundary condition in \fconsa\ is direct consequence of the trace formula
(2.14).

Note that this construction is pure algebraic, so, to justify it we encounter
no problems
typical for the bosonic case. Therefore, without assuming any implicit
condition we have by
construction that path integral \fz\ does not depend on $\alpha$, while FDA
is independent
of $\alpha$ up to $O(\epsilon)$. In Sect.5 we shall demonstrate this
explicitly.

Like in the bosonic case we can define the classical Hamiltonian of the
system
$H_{cl}(\xi^*, \xi)$ (see \ref\GT{D. M. Gitman, I. V. Tyutin, Quantization of
Fields with
Constraints (Springer-Verlag, 1990)},
\Zen and references therein), as $\lim_{\hbar \rightarrow 0}
H^{(\alpha)}(\xi^*, \xi)$. Then
$\alpha$-symbol of the Hamiltonian $\hat H$ coincides with the $H_{cl}$ if
the $\hat H$ is
$\alpha$-ordered image (3.23) of $H_{cl}$. But even in this case the exponent
in Eq.\fconsa\
is not any integral sum of the action of the corresponding classical
fermionic system, as the
exponent and the classical action are defined on different Grassmann algebras
(see \Zen\
for more detail).

Modification of Eq.\fconsa\ in the case of number of degrees of freedom more
then one is obvious.

\newsec{Non-local representation of FDA}

In this section we concentrate on the FDA for partition function in terms of
$\alpha$-symbols
and show that independently of the form of the Hamiltonian in both bosonic
and fermionic cases
the exponent of the FDA has a new representation which, except $\alpha = \pm
{1 \over 2}$, is
highly non-local.

\subsec{Bosonic system}

According to Eqs.\consg, \IIp, \qalha\ FDA for partition function in terms of
$\alpha$-symbols has the form
\eqn\consga{\eqalign{  Z^{(\alpha)}_N({t,t_0}) =
\int_{}^{}\prod_{ n\ =\ 1}^{ N}  d{q}_{n}{d{p}_{n} \over
2\pi} \;
\e{i \sum_{i = 1}^N [p_{i}\Delta q_{i} - \epsilon {H}_i^{({\alpha})}( p_i ,
q^{(\alpha)}(q_i, q_{i-1})]}  |_{q_0 = q_N},\cr} }
where $q^{(\alpha)}(q_i, q_{i-1})$ is given by Eq.\qalha. Thus, except the
case of $\alpha =
\pm {1 \over 2}$ the Hamiltonian part in \consga\ involves two $q^,$s of
neighbouring points.

Let us transform \consga\ in such a way that the Hamiltonian part in the
exponent has the form
${H}^{({\alpha})}( p_i , q_i)$.

Rewrite first \consga\ in terms of more convenient matrix notations.
Introduce
vectors ${ q}^T\equiv\left( {q_1,\cdots ,q_N} \right)$, ${ p}^T\equiv\left(
{p_1,\cdots
,p_N} \right)$, and
${ q}^{( \alpha
)T}$ $\equiv$ $( q^{( \alpha
)}(q_1, q_N)$, $\cdots$, $q^{( \alpha  )}(q_N, q_{N-1}))
$ ($T$ denotes the transpose) and
$N \times N$ matrices $\Delta_B$ and $A_{B}{\left( \alpha  \right)}$ such
that
\eqn\consg{
\left( {\Delta_B { q}} \right)^T=\left( {q_1-q_N,q_2-q_1,\cdots ,q_N-q_{N-1}}
\right),\quad { q}^{\left( \alpha  \right)}=A_{B}{\left( \alpha  \right)}{
q}.
}
Explicit form of $\Delta$ is obvious, while that of $A_{B}{\left( \alpha
\right)}$
is
\eqn\consh{
A_{B}{\left( \alpha  \right)} = \left( {\matrix{a&0&\cdots &0&b\cr b&a&\ddots
&\vdots
&0\cr 0&b&a&0&\vdots \cr
\vdots &\ddots &\ddots &\ddots &0\cr 0&\cdots &0&b&a\cr }} \right)\;,\quad
\left\{
\matrix{\displaystyle{a \equiv {1 \over 2}-\alpha,} \hfill\cr
\cr
\displaystyle{b \equiv {1 \over 2}+\alpha.} \hfill\cr} \right.
}
Then FDA \consga\ reads as
\eqn\consgb{\eqalign{
\int_{}^{}\prod_{ n\ =\ 1}^{ N}  d{q}_{n}{d{p}_{n} \over
2\pi} \;
\e{i { p}^T\Delta_B { q} - i\epsilon \sum_{i = 1}^N
{H}^{({\alpha})}( p_i , q^{(\alpha)}_i)}.\cr} }
Determinant of maurix $A_{B}{\left( \alpha  \right)}$
\eqn\consi{
\det A_{B}{\left( \alpha  \right)}=a^N-\left( {-b} \right)^N
}
does not vanish unless $\alpha = 0$, $N$ is even. Therefore,
excluding this case we can make in \consgb\ change of variables
${ q}' = { q}^{\left( \alpha  \right)}$,
which leads it to the form
\eqn\consgc{\eqalign{
{1 \over \det A_{B}{\left( \alpha  \right)}} \int_{}^{}\prod_{ n\ =\ 1}^{ N}
d{q}_{n}{d{p}_{n}
\over 2\pi} \;
\e{i { p}^T \Delta_B(\alpha) { q} - i\epsilon \sum_{i = 1}^N
{H}_i^{({\alpha})}( p_i , q_i)},\cr} }
where we have omitted the primes and denoted
\eqn\kin{ \Delta_B(\alpha) \equiv \Delta_B A_{B}^{-1}{\left( \alpha
\right)}.}
This is exactly what we wanted, and all what is to be done is to find an
explicit form of
matrix $\Delta_B(\alpha)$.

Note that $A_{B}\left(\alpha \right)$ is represented in the form
\eqn\consk{
A_{B}\left( \alpha  \right)=a I +b \Omega_{B},\quad \Omega_B^N=I,
}
where $I$ is $N \times N$ unit matrix. Then using identity
\eqn\consl{
\left( {1-c^N} \right)I=\left( {I-c\Omega _B}
\right)\sum\limits_{k=0}^{N-1} {\left( {c\Omega _B} \right)^k},
}
where $c$ is a parameter, we obtain
\eqn\consm{
A_{B}^{-1}\left( \alpha  \right)={1 \over {a^N-\left( {-1}
\right)^Nb^N}}\sum\limits_{k=0}^{N-1} {\left( {-1}
\right)^ka^{N-1-k}b^k\Omega
_B^k}.
}
Since matrix $\Delta_B$ in \consgc\ is also expressed in terms of
$\Omega _B$ as $\Delta_B=I-\Omega _{B}$, we have
\eqnn\consn
$$\eqalignno{
\Delta_B(\alpha) = & {1 \over {a^N-\left( {-1}\right)^Nb^N}} & \consn \cr
\times &\left[ {\left( {a^{N-1}-\left( {-1} \right)^{N-1}b^{N-1}}
\right) I + \sum\limits_{k=1}^{N-1} {\left( {-1}
\right)^ka^{N-1-k}b^{k-1}\Omega _B^k}} \right]. &
}
$$
In terms of step function
\eqn\conso{
\theta \left( {i-j} \right)=\left\{ \matrix{1,\quad  { i\ge j},
\hfill\cr
\cr
0,\quad  { i<j}, \hfill\cr} \right.
}
matrix elements of $\Omega _B^k$ are given by
\eqn\consp{
\left( {\Omega _B^k} \right)_{ij}=\delta _{i,j+k}\theta \left( {N-k-j}
\right)+\delta _{i,j+k-N}\theta \left( {j+k-N-1} \right),
}
so that explicit form of matrix $\Delta_{B}\left( \alpha  \right)$ reads as
\eqn\consq{
\eqalign{\left[ {\Delta_{B}\left( \alpha  \right)} \right]_{ij}=&{1 \over
{\left( {\displaystyle\alpha +{1\over 2}} \right)^N-\left(
{\displaystyle\alpha
-{1\over 2}} \right)^N}}\left\{ {\left[ -{\left( {\alpha +{1\over 2}}
\right)^{N-1}+\left( {\alpha -{1\over 2}} \right)^{N-1}}
\right]\delta _{i,j}} \right.\cr
  &+\theta \left( {i-j-1} \right)\left(
{\alpha +{1\over 2}} \right)^{i-j-1} \left( {\alpha -{1\over 2}}
\right)^{N-1-i+j}\cr
  &\left. {+\theta \left( {j-i-1} \right)
\left( {\alpha +{1\over 2}} \right)^{N-1+i-j}\left( {\alpha -{1\over 2}}
\right)^{j-i-1}} \right\}.\cr}
}

Thus, we see that only in the cases of $\alpha = \pm {1 \over 2}$ ($pq$- and
$qp$-orderings),
when we have
\eqn\pqp{\eqalign{& [\Delta_{B}({1 \over 2})]_{ij} = \delta_{i+1, j} -
\delta_{i,
j} \quad (i < N),
\cr & [\Delta_{B}(-{1 \over 2})]_{ij} = \delta_{i, j} - \delta_{i-1, j} \quad
(i > 1),\cr}}
expression $\Delta_{B}({\alpha}) q$ corresponds to (right and left)
differences
approximating time derivative in the term $p {\dot q}$ in the exponent.
Any other values of $\alpha$ determine the highly non-local form of the
kinetic part which have
no continuum limit and which cannot be regarded as an approximations for the
classical
term $p {\dot q}$. Note also, that matrix $\Delta_{B}\left( \alpha  \right)$
is
antisymmetrical only for value $\alpha = 0$ corresponding to the Weyl
ordering (remind that
for $\alpha = 0$ this representation exists only for $N$ is odd).

\subsec{Fermionic system}

In this case our program is the same: transforming FDA \fconsa\ in such a way
that
its Hamiltonian part has the form $H^{(\alpha)}(\xi^*_i, \xi_i)$. In view of
the similarity of
the structure of the FDA in fermionic and bosonic cases, all the steps are
also
essentially the same.

With similar matrix notations we can rewrite \fconsa\ as
\eqn\fconsb{
\eqalign{Z^{(\alpha)}_N(t, t_0)
 = \int \prod\limits_{n=1}^{N} d\xi_n d\xi^*_n \; \e{- \xi^\dagger \Delta_F
\xi
 - i \epsilon \sum_{i=1}^N H^{(\alpha  )}(\xi_i^*, \xi^{(\alpha)}_i)}, \cr} }
where
\eqn\fconsc{
\left( {\Delta_F {\xi}} \right)^T=\left( {\xi_1+\xi_N,\xi_2-\xi_1,\cdots
,\xi_N-\xi_{N-1}}
\right),\quad {\xi}^{\left( \alpha  \right)}=A_{F}\left( \alpha  \right)
{\xi},
}
and
\eqn\fconsd{ A_{F}\left( \alpha  \right)=\left( {\matrix{a&0&\cdots &0&-b\cr
b&a&\ddots &\vdots &0\cr 0&b&a&0&\vdots \cr
\vdots &\ddots &\ddots &\ddots &0\cr 0&\cdots &0&b&a\cr }} \right)\;,\quad
\left\{
\matrix{\displaystyle{a \equiv {1 \over 2}-\alpha}, \hfill\cr
\cr
\displaystyle{b \equiv {1 \over 2}+\alpha}. \hfill\cr} \right.
}
The principal difference of these expressions from those of \consg, \consh:
signs of matrix elements $(\Delta_F)_{1N}$ and $[A_F(\alpha)]_{1N}$, comes
from antiperiodicity
of boundary conditions in \fconsa.
Determinant of $A_F(\alpha)$
\eqn\fconse{
\det A_{F}\left( \alpha  \right)=a^N+\left( {-b} \right)^N
}
does not vanish unless $\alpha = 0$, $N$ is odd (cf., \consi). Then,
excluding this
case and making the change of variables we come to the desirable expression
\eqn\fconsd{
\eqalign{Z^{(\alpha)}_N(t, t_0)
 = \det A_F(\alpha) \int \prod\limits_{n=1}^{N} d\xi_n d\xi^*_n \; \e{-
\xi^\dagger
\Delta_F(\alpha)
\xi
 - i \epsilon \sum_{i=1}^N H^{(\alpha  )}(\xi_i^*, \xi_i)}, \cr} }
with
\eqn\kinf{ \Delta_F(\alpha) \equiv \Delta_F A_{F}^{-1}{\left( \alpha
\right)}.}

In the way completely analogous to the one of the previous section we find
\eqn\fconsk{
\eqalign{\left[ {\\elta_{F}\left( \alpha  \right)} \right]_{i j}=&{-1
\over {\left( {\displaystyle\alpha +{1\over 2}} \right)^N+\left(
{\displaystyle\alpha
-{1\over 2}} \right)^N}}\left\{ {\left[ {\left( {\alpha +{1\over 2}}
\right)^{N-1}+\left( {\alpha -{1\over 2}} \right)^{N-1}}
\right]\delta _{i,j}} \right.\cr
  &+\theta \left( {i-j-1} \right)\left(
{\alpha +{1\over 2}} \right)^{i-j-1} \left( {\alpha -{1\over 2}}
\right)^{N-1-i+j}\cr
  &\left. {-\theta \left( {j-i-1} \right)\left( {\alpha +{1\over 2}}
\right)^{N-1+i-j}
\left( {\alpha -{1\over 2}}
\right)^{j-i-1}} \right\}.\cr}
}

Like in bosonic case the kinetic part of the exponent in \fconsb\ can be
interpreted as
fermion analogue of $p \dot q$ only for the same two values of $\alpha$:
\eqn\pqpf{\eqalign{& [\Delta_{F}({1 \over 2})]_{ij} = \delta_{i+1, j} -
\delta_{i,
j} \quad (i < N),
\cr & [\Delta_{F}(-{1 \over 2})]_{ij} = \delta_{i, j} - \delta_{i-1, j} \quad
(i > 1).\cr}}
For any other $\alpha$ the expression is non-local. In particular, for
$\alpha = 0$
(the Weyl ordering) we have
\eqn\Wf{[\Delta_{F}(0)]_{i j} = 2 \left[\sum_{l = 1}^{N - i} (-1)^{l+1}
\delta_{i + l, j} -
\sum_{l = 1}^{i - 1} (-1)^{l+1} \delta_{i - l, j}\right]}
(contrary to
the bosonic case this representation exists only for $.$ is even). In this
case the
expression is maximally non-local in the sense that matrix elements
$\Delta_{F}(\alpha)_{i
j}$ do not decrease at all with increasing $|i - j|$, matrix
$\Delta_{F}(\alpha)$ being antisymmetrical.

Representations \fconsd\ for particular cases \pqpf\ and \Wf\ have been
obtained in \Ze\
using the above mentioned technique of $*$-product for the symbols of
operators, passing more familiar representation \fconsb.

The nontrivial point now is that despite the drastical change of the form of
the
exponent of the FDA with change of $\alpha$, the FDA itself, that is $Z_N(t -
t_0)$, is
independent of the $\alpha$ up to terms $O(\epsilon)$. To demonstrate this
explicitly we shall
consider a simple

\subsec{Example}

Namely, consider the fermionic system governed by the Hamiltonian
$\hat{H}= \omega \hat{b}^\dagger\hat{b}$ (fermionic oscillator) in which case
the partition
function is extremely simple:
\eqn\pff{Z(t - t_0) = 1 + \e{- i\omega (t - t_0)}.}
In view of \IIbff, FDA \fconsd\ for the partition function reads as
\eqn\fconsdd{
\eqalign{Z^{(\alpha)}_N(t, t_0)
 = \det A_F(\alpha) \int \prod\limits_{n=1}^{N} d\xi_n d\xi^*_n \; \e{-
\xi^\dagger
[\Delta_F(\alpha) + i \epsilon \omega ]
\xi - i ({1 \over 2} - \alpha) \omega (t - t_0)}. \cr}
}
The integration is easy, so we have
\eqn\znf{\eqalign{Z^{(\alpha)}_N & (t, t_0) =  \det \left[- A_F(\alpha)
\Delta_F(\alpha) - i
\epsilon
\omega A_F(\alpha) \right] \e{- i ({1 \over 2} - \alpha) \omega (t - t_0)}
\cr
& =\left[ \left(1 + i \epsilon ({1 \over 2} - \alpha ) \omega \right)^N +
\left(1 - i \epsilon ({1 \over 2} + \alpha ) \omega \right)^N \right]
\e{- i ({1 \over 2} - \alpha) \omega (t - t_0)}, \cr}}
from where it follows explicitly that
\eqn\eps{Z^{(\alpha)}_N(t, t_0) = Z(t - t_0) + O(\epsilon, \alpha).}

We can get a more detailed information about the intrinsic dependence of the
FDA on $\alpha$ if
we use for each infinitesimal time interval different $\alpha$-symbols, that
is
introduce different $\alpha^,$s for different terms of the product in
\fconsa. Modification of
expression \fconsd\ in this case is obvious. Consider now quantity $\partial
Z_{N}^{(\alpha)}/
\partial \alpha_{i}$. Then, after straightforward calculations we get
\eqn\indepi{
{\partial \over \partial \alpha_{i}}Z_{N}^{(\alpha)}= - (\epsilon \omega
)^{2}
\left[ \left(\Delta_F\left( \alpha
\right) + i \epsilon \omega \right)^{-1}\right]_{i i}Z_{N}^{(\alpha)}.
}
This result explicitly displays that $\alpha$-dependence of each term forming
the
FDA is of order $O(\epsilon^2)$, in its turn, demonstrating the
self-consistency of all our
construction.

\newsec{Summary and discussion}

To summarize our results, we obtained the general form of FDA for bosonic
\consf, \consg\ and
fermionic \fconsa\ systems in terms of symbols of operators defined in
Sect.3; argued that the
FDA are independent of types of the symbols up to $O(\epsilon)$ (Sects.4,
5.3); found new,
non-local, representations for the FDA \consgc, \consq, and \fconsd, \fconsk.

In quantum mechanics the choice of of type of the symbols as well as of the
representation of
the FDA, can only be dictated by reasons of convenience or naturalness.
Indeed, the discovered
non-locality in the exponent of FDA is a property of the FDA itself, rather
than underlying
quantum system. This however can become a dynamical problem if additional
symmetries like a
gauge invariance are imposed.

This is also the case when FDA are used as a basis for formulation of the
Euclidean quantum
field theory on a lattice (in this case $\epsilon$ plays a role of the
lattice spacing). Then,
if corresponding lattice subgroup of Euclidean group is required to be the
symmetry of the
theory at the regularization level, the form of the kinetic part of the FDA
certainly
affects the form of its Hamiltonian part. This has been used for defining
fermion actions on a
lattice \ref\KS{T. Kashiwa, H. So, Prog. Theor. Phys. {\bf 73}, 762 (1985)},
\Ze, where
additional requirement of chiral invariance leads to a non-local formulation
of the Weyl
fermions on a lattice \Ze. It is using the explicit form of FDA that
guarantees a lattice
theory to be free of such pathologies as ``species doubling".

\bigbreak\bigskip
\centerline{{\bf Acknowledgments}}
\vskip 8pt

The work of S.V.Z. was supported by JSPS; he is grateful to the Elementary
Particle Theory
Group of Kyushu University for the warm hospitality.

\listrefs

\end